# Thin Film Synthesis, Structural Analysis, and Magnetic Properties of Novel Ternary Transition Metal Nitride MnCoN$_2$

Sita Dugu[1*], Rebecca W. Smaha[1], Shaham Quadir[1], Andrew Treglia[2], Shaun O'Donnell[1,2], Julia Martin[1], Sharad Mahatara[1], Glenn Teeter[1], Stephan Lany[1], James R. Neilson[2,3], Sage R. Bauers[1†]

[1] *Materials Sciences Center, National Renewable Energy Laboratory, 15013 Denver West Parkway, Golden, Colorado 80401, United States.*
[2] *Department of Chemistry, Colorado State, University, Fort Collins, Colorado 80523-1872, United States.*
[3] *School of Advanced Materials Discovery, Colorado State University, Fort Collins, Colorado 80523-1872, United States.*

\* sita.dugu@nrel.gov
† sage.bauers@nrel.gov

1. Abstract

Recent high-throughput computational searches have predicted many novel ternary nitride compounds providing new opportunities for materials discovery in underexplored phase spaces. Nevertheless, there are hardly any predictions and/or syntheses that incorporate only transition metals into new ternary nitrides. Here, we report on the synthesis, structure, and properties of MnCoN$_2$, a new ternary nitride material comprising only transition metals and N. We find that crystalline MnCoN$_2$ can be stabilized over its competing binaries, and over a tendency of this system to become amorphous, by controlling growth temperature within a narrow window slightly above ambient condition. We find that single-phase MnCoN$_2$ thin films form in a cation-disordered rocksalt crystal structure, which is supported by ab-initio calculations. X-ray photoelectron spectroscopy analysis suggests that MnCoN$_2$ is sensitive to oxygen through various oxides and hydroxides binding to cobalt on the surface. X-ray absorption spectroscopy is used to verify that Mn$^{3+}$ and Co$^{3+}$ cations exist in an octahedrally-coordinated environment, which is distinct from a combination of CoN and MnN binaries and in agreement with the rocksalt-based crystal structure prediction. Magnetic measurements suggest that MnCoN$_2$ has a canted antiferromagnetic ground state below 10 K. We extract a Weiss temperature of $\theta$ = –49.7 K, highlighting the antiferromagnetic correlations in MnCoN$_2$.

## 2. Introduction

Transition metal (*TM*) nitrides are important in several industrial applications such as superconductors, hard coatings, plasmonics, photovoltaics, piezoelectrics, permanent magnets, and more.[1–4] Despite their known and promising functionality, *TM* nitride materials have been historically underexplored due to challenging synthesis routes and hurdles in achieving oxygen-free pure nitrides. To expand the number of known functional nitrides, several papers over the last decade have used high-throughput computations to identity nitride materials for various applications including earth-abundant semiconductors, magnetic antiperovskites, and MAX phase precursors to MXenes.[5–7] The bounded nature of these works resulted in a few dozen nitride compounds to study, many of which were known. On the other hand, in 2019 Sun et al[8] conducted a properties- and structure-agnostic high-throughput computational search for ternary nitride compounds using a data-mined structure predictor[9] algorithm specifically trained for new nitride discovery. This search found 244 newly predicted stable ternary nitride phases, out of which 93 were in completely new chemical spaces (i.e., a pair of metals previously unknown to form a ternary nitride). Guided by these predictions, experimentalists have recently utilized radio frequency (rf) sputtering to synthesize several new ternary nitrides in the laboratory such as $MgTiN_2$, $MgHfN_2$, $Mg_2NbN_3$,[10] $MgZrN_2$,[11] $ZnTiN_2$,[12] $ZnZrN_2$,[13] $Zn_2SbN_3$,[14] $LaWN_3$,[3] $MnSnN_2$,[15] $Zn_2TaN_3$,[16] and $Zn_2VN_3$,[17] to name a few.

While this is a fruitful materials discovery approach, none of these compounds contain only transition metals (*TM*s) (here, we consider Zn as chemically more like an alkaline earth, *AE*, main group element than a transition metal). In fact, the 2019 paper[8] only found one new *TM*-*TM*-N chemical space containing a stable ternary nitride, Mn–Co–N. **Figure 1a** summarizes the new and known *AE-TM*-N and *TM-TM*-N as reported by this paper. Because many of the known *TM*-N binary compounds are useful compounds and magnetic materials, the predicted $MnCoN_2$ phase is a compelling materials discovery target. The structure predicted for $MnCoN_2$ in 2019 (entry mp-1029367 in the Materials Project)[18] is in a trigonal space group (sg, $R3m$) where the cations (Mn and Co) are bonded with anionic N to form $MnN_4$ and $CoN_4$ tetrahedra. Unlike the chalcopyrite structure, the predicted R3m structure contains two inequivalent $N^{3-}$ sites. On the first $N^{3-}$ site, $N^{3-}$ is bonded to one Mn cation and three equivalent Co cations to form $NMnCo_3$ tetrahedra. In the second $N^{3-}$ site, $N^{3-}$ is bonded to one Co cation and three equivalent cations of Mn to form corner sharing $NMn_3Co$ tetrahedra.

It is important to contextualize $MnCoN_2$ among the competing $MnN_x$ and $CoN_x$ binary phases. While there are several in each system, we focus on those with a 1:1 metal:N ratio, like the $MnCoN_2$ compound of interest. We begin by discussing manganese nitrides. The manganese nitride with 1:1 cation:anion composition, θ-MnN, exists experimentally in a tetragonally distorted rock salt (RS) structure (space group number 139) with antiferromagnetic (AFM) order with $T_N$ = 650 K, measured from an exfoliated powder originally prepared by magnetron sputtering.[19]

Conversely, some theories have proposed a cubic zinc-blende (ZB) structure under certain conditions[20,21]. Convex hull calculations from Li et al.[21] predict a phase transition sequence for MnN with increasing pressure; first the semiconducting non-magnetic ZB phase transforms at 5 GPa to a metallic AFM phase in the NiAs structure, which at 40 GPa is destabilized vs. a metallic ferromagnetic (FM) RS phase. Interestingly, while the low-pressure phases are not seen experimentally, on increasing pressure, Zheng *et al.*[22] recently observed a transition from AFM to FM order in rocksalt-derived MnN at 34 GPa. The transition to FM was associated with a reduction in cell volume of ca. 12% to 16Å$^3$.

There are also several known nitrides of cobalt. While most of the early 1:1 *TM*:nitrides are reported as RS, this structure becomes destabilized with increasing electron count from populating antibonding orbitals.[23] Depending on the synthesis method, CoN had been reported as RS[24] and ZB.[25] Because both RS and ZB have underlying face-centered-cubic (FCC) lattice symmetry, the same family of x-ray diffraction (XRD) peaks are present in both. However, relative intensities in powder XRD patterns and peak positions both change between polymorphs. Suzuki et al.[26] synthesized ZB CoN by rf sputtering in 1995. In lieu of a full refinement, they showed the material is ZB by comparing experimental XRD intensity with simulated patterns of CoN in both ZB and RS structures. They further observe that the ZB CoN exhibits Pauli paramagnetic properties which was later theoretically confirmed by Lukashev et al[27] using spin polarized calculations.

Considering the *TM*–N binaries, the unreported MnCoN$_2$ compound must be stabilized among multiple competing FCC structures, namely tetragonal RS MnN with octahedral coordination and cubic ZB CoN with tetrahedral coordination. Thus, the structure of MnCoN$_2$ might be either one of these or a mixture of both. In this work, we study this novel phase space and report the experimental synthesis of MnCoN$_2$, which adopts a cation-disordered rocksalt structure. We use laboratory and synchrotron XRD data, along with first principles calculations, to explore the crystal structure, comparing the experimental data to the binary structures and the previously predicted structure. The octahedral environment is further supported by XAS experiments. We also show that MnCoN$_2$ exhibits canted AFM order at low temperature. Altogether, this work should motivate the identification and study of additional new *TM$_1$*-*TM$_2$*-N materials.

3. RESULTS AND DISCUSSION

   3.1. Computational prediction:

Since a high-throughput structure predictor approach does not always identify ground state structures,[8] we begin by more thoroughly surveying candidate configurations for MnCoN$_2$. We selected five structural prototypes for MnCoN$_2$ and considered different possible magnetic configurations within a small eight atom cell. The prototypes include three zinc-blende (ZB) derived structures: the MnCoN$_2$ phase from the Materials Project (mp-1029367, sg-160, non-magnetic), chalcopyrite ordered ZB (sg-122), and CuAu ordered ZB (sg-115), as well as two rocksalt (RS) derived structures from our previous work on similar ternary nitrides:[10] α-NaFeO$_2$ ordered RS (sg-13) and γ-LiFeO$_2$ ordered RS (sg-141). **Figure 1b** compares the total energy of the prototypes considered in their most stable magnetic configurations. The lowest total energy was found for the α-NaFeO$_2$ type (sg-13) structure in an AFM configuration (zero reference energy as shown in **Figure 1b**), in which both the Mn and Co substructures have AFM order with local moments of ~3.3 μ$_B$ and ~2.5 μ$_B$, respectively. These moments suggest high-spin Mn$^{4+}$ and Co$^{2+}$ oxidation states. This AFM structure is ~18 meV/atom lower in energy than the corresponding ferromagnetic (FM) configuration. We also tested ferrimagnetic (with Mn and Co having opposite spin direction) and low-spin configurations for Co, which were found to be even less stable than FM.

The second most stable prototype structure, with a polymorph energy difference of 36 meV/atom, is a RS (sg-141) structure with an AFM high spin Mn (~3.7 μ$_B$) and low-spin Co (~0.3 μ$_B$) magnetic configuration. The sg-13 and sg-141 structures differ in their cation ordering, but the moderate energy difference suggests that thin-films could be stabilized in a cation-disordered rocksalt structure[10]. The calculated equivalent cubic lattice constants of the RS structures are 4.15 Å for sg-13 and 4.09 Å for sg-141.

The ZB-derived structures were all calculated assuming ferrimagnetic ordering with Mn (~3.1-3.4 μ$_B$) and Co (~2.3 - 2.6 μ$_B$) spins in opposite directions. Also, there is a significant local moment of ~0.3 μ$_B$ on nitrogen, suggesting partial reduction of N to N$^{3-}$ and ambiguity in the oxidation states of Mn and Co. The lowest energy is obtained for chalcopyrite (sg-122) at 97 meV/atom above the α-NaFeO$_2$ type ground state. The non-magnetic MnCoN$_2$ (sg-160 ZB) structure is energetically very unstable at an energy of 538 meV/atom. Even in the ferrimagnetic state, it is still the least stable of the three ZB structures at 149 meV/atom. All three ZB structures have equivalent cubic lattice constants around 4.42 Å in the ferrimagnetic configuration, considerably larger than and well distinguishable from the RS structures. However, the lattice constant of the (sg-160 ZB) structure is (artificially) reduced to 4.22 Å in the unstable non-magnetic configuration, which is comparable to the lattice parameters of the stable RS structures. Structure schematics of lowest energy level RS-derived (sg-13) and ZB-derived (sg-122) phases are shown in right side of **Figure 2b.**

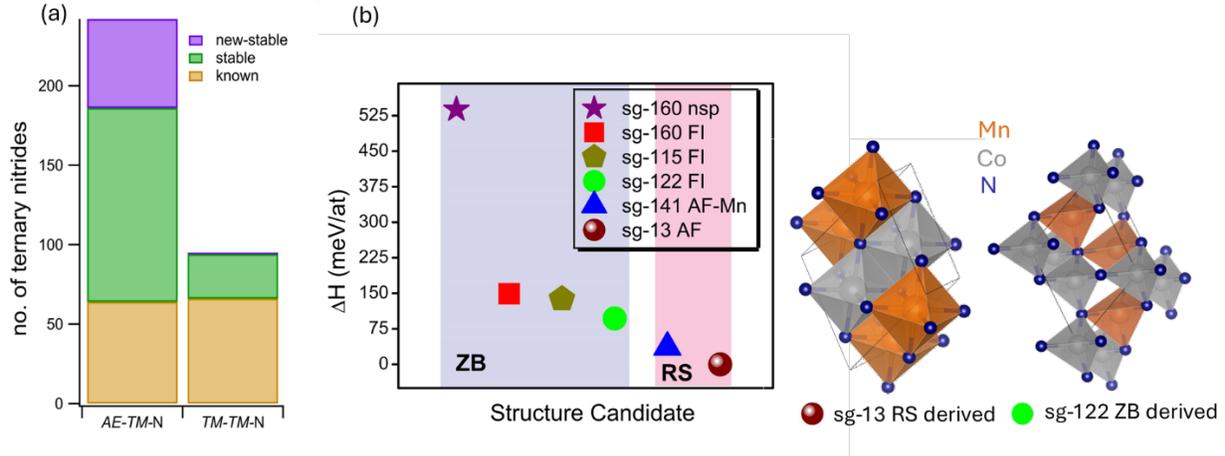

Figure 1: (a) Bar chart of known, predicted metastable and new stable nitrides of alkaline earth –transition metal nitrides (*AE-TM*-N) and transition metal – transition metal nitrides (*TM-TM*-N). (b) Calculated enthalpies of five prototype structural candidates, sg-13 RS-derived and sg-122 ZB-derived crystal structure shown as the right side of (b)

3.2. MnCoN$_2$ synthesis

**Figure 2a** comprises 44 laboratory XRD patterns collected as a function of position on a compositionally graded Mn-Co-N film on a silicon substrate. The diffraction intensity is presented as a heatmap against the measured cation composition and scattering vector, $Q$ ($\lambda$ = 1.54059 Å). The film was grown without active heating, which we refer to as "ambient" in this article. The traces on the bottom of the plot are simulated XRD patterns; the black curve is the reference pattern for ZB CoN[26], whereas the gray curve is that of tetragonally distorted RS MnN[19]. The characteristic FCC peaks from the ZB simulation are shifted to lower $Q$ values than RS, which is expected from the relatively less dense ZB structure. Two phases are observed for samples deposited at ambient temperature: ZB at lower Mn concentrations and RS at higher Mn content, which is consistent with the ZB phase of CoN[26] and RS phase of MnN.[19] Similar behavior is seen with active heating with film deposition temperature ($T_d$) at 50 °C, but the RS region shifts to lower Mn compositions, even beyond the stoichiometric 1:1 Mn:Co condition. Samples grown at 150 °C exhibit a single phase for all Mn concentrations studied, as shown in **Figure S1a**. Despite observation of a single RS phase at high $T_d$, the overall crystallinity of the film is lower as indicated by diffraction peaks with large full width at half maximum (FWHM), a trend which increases with higher deposition temperatures. The FWHM of the (111) and (200) peaks for $T_d$ of ambient, 50 °C and 150 °C are shown in **Figure S1b**.

Films grown at several conditions are summarized on a phase diagram as a function of deposition temperature ($T_d$) and composition in **Figure 2b**, where each plot marker color represents a different compositionally graded film. Films grown at 150 °C and above only show a RS phase regardless of Mn/(Mn+Co) values. However, the crystallinity of the samples progressively decreases with an increase in deposition temperature, highlighted by the color gradient area in

**Figure 2b**. For samples deposited at ambient and up to 50 °C, the observed phases (ZB and RS in our system) depend on the Mn/(Mn+Co) values. For films grown at 50 °C, the ZB phase is observed when Mn/(Mn+Co) < 0.34 and RS is observed when Mn/(Mn+Co) > 0.45; both phases appear in the range $0.34 \leq$ Mn/(Mn+Co) $\leq 0.45$. At ambient, these stability windows are moved to higher Mn content; phase pure ZB is seen when Mn/(Mn+Co) < 0.38 and phase-pure RS when Mn/(Mn+Co) > 0.52 with mixed ZB and RS at intermediate compositions.

To further diagnose the crystallinity of samples, we characterized them with synchrotron grazing incidence wide angle X-ray scattering (GIWAXS; $\lambda = 0.97625$ Å). **Figure 2c** shows integrated GIWAXS patterns of films grown at ambient with Mn/(Mn+Co) = 0.4, 0.5, and 0.6. For the compositions Mn/(Mn+Co) = 0.5 and 0.4, peak splitting is observed indicative of RS/ZB coexistence; Mn/(Mn+Co) = 0.6 is closest to phase pure. To understand the role of thermal conductance of the substrate on phase formation, films were subsequently also grown on borosilicate glass (Corning Eagle XG glass (EXG)) at ambient conditions. Despite the lack of active heating, both the lower thermal conductivity (<1 $Wm^{-1}K^{-1}$ for glass vs >100 $Wm^{-1}K^{-1}$ for Si) and thicker substrate (1.1mm for glass vs. 0.6mm for Si), coupled with front-side heating from the plasma, result in an elevated growth temperature during deposition on glass. GIWAXS patterns from the same Mn/(Mn+Co) = 0.4, 0.5, and 0.6 compositions, but grown on glass substrates are shown in **Figure 2d**. With the slightly elevated deposition temperature afforded by deposition onto the glass substrate, phase-pure RS is achieved for Mn/(Mn+Co) = 0.5 and 0.6; only at high Co content is a mixed phase observed. The detector images of films grown on Si substrate and EXG are shown in **Figure S2**.

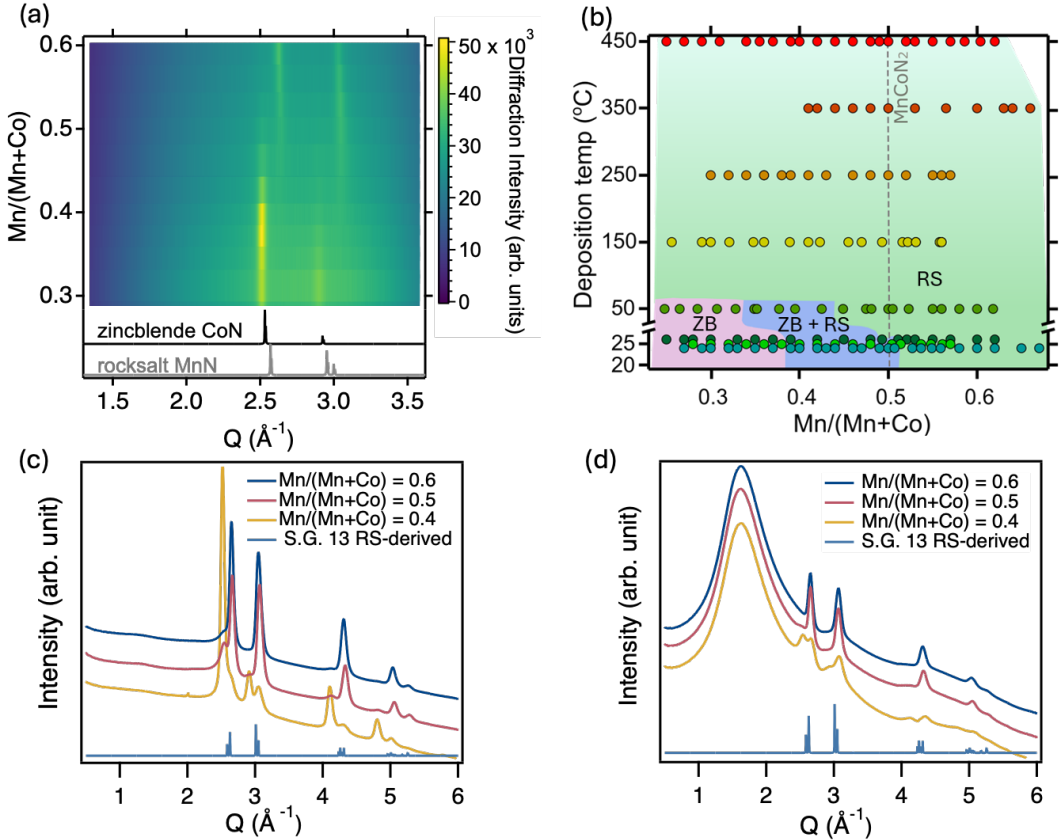

Figure 2: (a) Laboratory XRD heatmap of a MnCoN$_2$ film grown at ambient conditions. Reference diffraction patterns of ZB CoN (black curve, ICSD entry #79936) and tetragonally distorted RS MnN (gray curve, ICSD entry #106932) are placed at the bottom. (b) Experimental phase map of MnCoN$_2$ deposited at various temperatures with all compositions measured by XRF. The shaded pink area indicates ZB phase while the green area represents RS phase. The blue shaded area represents the region where both RS and ZB region are observed. The light green gradient with increasing temperature indicates lower crystallinity of the RS phase. Points with the same plot marker color were grown as part of the same combinatorial deposition. (c) and (d) Synchrotron GIWAXS data measured from three different Mn/(Mn+Co) compositions for films grown on (c) pSi and (d) EXG glass. Both films were grown in nominally ambient conditions. The calculated pattern in space group 13 is placed at the bottom.

To summarize **Figure 2**, the apparent phase instability from minor thermal differences created by the plasma environment or different substrates during ambient temperature deposition suggests the growth process is sensitive to the thermal environment. At the same time, deposition temperatures ≥150°C reduce crystallinity. One explanation for this could be N loss at elevated $T_d$; MnCoN$_2$ lacks an electropositive metal that helps stabilize N-rich nitrides,[8] and instead contains only transition metals well known to accommodate substoichiometric N.[28] For films grown just above ambient (i.e., at 50 °C), the metal-stoichiometric films are highly crystalline and reliably form in the RS structure.

### 3.3. Chemistry of Mn and Co in MnCoN$_2$

X-ray photoelectron spectroscopy (XPS) is a powerful tool to understand the electronic and surface chemical environment of a system. XPS was conducted at 3-4 points with varying Mn:Co concentration for films grown at 50 °C and 150 °C on a Si substrate. **Figure 3 (a-d)** shows the Mn 2p, Co 2p, N 1s, O 1s spectra acquired from the MnCoN$_2$ film grown at 50 °C. The Co and Mn 2p orbitals (**Figure 3a and 3c**) each display characteristic spin-orbit splitting (2p$_{3/2}$ versus 2p$_{1/2}$). Both the Mn 2p and N 1s spectra (**Figure 3a and 3b**) can be fit with asymmetric peaks consistent with a single chemical state for each element. In the Co 2p core level, the two main peaks at 780 eV and 795.4 eV correspond to Co 2p$_{3/2}$ and Co 2p$_{1/2}$ respectively.[29,30] The core levels (**Figure 3c**) require peaks corresponding to at least three distinct chemical states, with the presence of additional shakeup peaks and a contribution from the Mn 2s orbitals. The dark blue (780.15 eV and 795.55 eV) and light blue (782.2 eV and 797.4 eV) peaks are tentatively associated with Co oxides and hydroxides.[31,32] The presence of hydroxides is further supported by the appearance of several shakeup peaks, observed at 785.9 eV, 801.4 eV, 790.5 eV, and 805.8 eV which are often associated with the adsorption of surface oxides/hydroxides.[33,32] The peak at 771.8 eV is attributed to Mn 2s, while the peaks at 778.6 eV and 794.15 eV are assigned to metallic Co. Collectively, these results indicate that Co–O bonds form preferentially in the near-surface region of MnCoN$_2$ during air exposure, leading to decomposition into Co-(O,OH) and Mn-N phases.

XPS indicates the presence of large amounts of oxygen in the near-surface region, which is also seen in the bulk of a film by Rutherford backscattering spectrometry (RBS), shown in **Figure S3**. This suggests the presence of surface oxides, not unexpected for nitrides exposed to atmosphere.[33] O 1s spectra (**Figure 3d**) can be subdivided into two peaks at 531.2 eV and 529.5 eV. The blue-shaded O 1s feature centered at 529.5 eV is due to lattice O$^{2-}$, and the position is consistent with that expected for metal oxide(s).[34,35] The higher binding energy O 1s (purple shaded) featured at 531 eV likely corresponds to the surface hydroxyls, under coordinated lattice oxygen,[36] and/or traces of oxidized contaminants.[35,37] A small yellow shaded peak featured at higher binding energy (533.2 eV) is attributed to carbon contamination.

Similar XPS spectra, shown in **Figure S4 (a-d)**, were collected at multiple compositions in the range 0.4 < Mn/(Mn+Co) < 0.6, for deposition temperatures of 50 ºC and 150 ºC. The integrated intensity of the Mn, Co, and N peaks discussed above are plotted against each other in **Figure 3e**. This plot shows a strong linear correlation between the Mn 2p and N 1s peak areas, indicating that these constituents very likely participate in the same phase. Further evidence for the existence of a Mn-N phase is the linear correlation (with slope = 1) between binding energies for the Mn 2p and N 1s peak, shown in **Figure 3f**. At the same time, the sum of non-metallic Co 2p peak areas is anti-correlated with the N 1s peak area, suggesting that there are no phases present that involve both Co and N. Based on all of these observations, we assign the Mn 2p and N 1s peaks to a MnN$_x$ phase, as annotated in **Figure 3(a-b)**, with a Mn:N ratio ~1.7 based on the observed sensitivity-factor corrected peak areas. The ΔBE =244.64 ± 0.01 eV value represents the characteristic binding energy separation between the N 1s and Mn 2p peaks for the MnN$_x$ phase. On the other hand, the non-metallic Co 2p peaks appear anticorrelated to N 1s. In fact, the total peak area of Co 2p (Co -oxides and hydroxides) has a positive correlation with M – O, whereas Mn does not, shown in **Figure S5**. This analysis indicates that Mn–N bonds are likely stronger and better resist replacement of N with O compared to Co–N bonds on the surface of MnCoN$_2$, and that the nitrogen detected is associated with Mn.

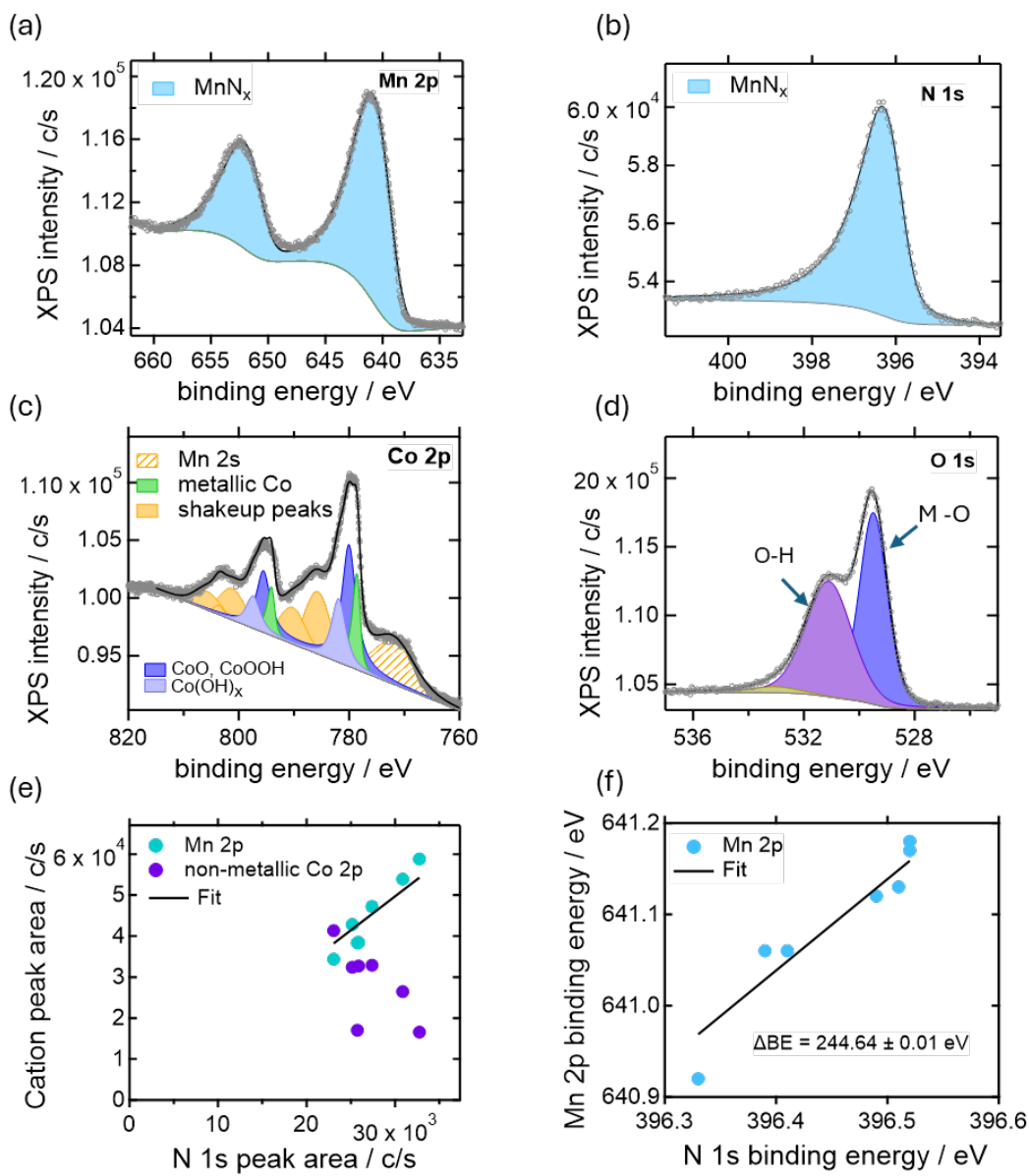

Figure3: Surface XPS spectra of MnCoN$_2$. (a-d) are XPS spectra of Mn2p, N1s, Co2p and O1s of stoichiometric MnCoN$_2$ grown at 50 ºC. (e) The cation peak area versus N 1s peak area for compositional variation 0.35 < Mn/(Mn+Co) < 0.58. (f) Binding energy of Mn 2p versus N 1s for compositional variation 0.35 < Mn/(Mn+Co) < 0.58, with a linear fit.

Chemical information of the films was further studied using X-ray absorption spectroscopy (XAS) to understand the bulk chemistry of the system. Three compositions with Mn/(Mn+Co) = 0.6, 0.5 and 0.4 for films grown at ambient, 50 °C and 150 °C were selected and analyzed for X-ray absorption near-edge structure (XANES) spectroscopy and extended X-ray absorption fine structure (EXAFS) spectroscopy on both the Mn and Co K-edges. The Mn K-edge XANES of films grown at ambient for Mn/(Mn+Co) value of 0.6, 0.5, and 0.4 is presented in **Figure 4a**, along with references MnO and $Mn_2O_3$ for comparison.

The absorption edges ($E_o$) for all three Mn/(Mn+Co) compositions are in the range 6547 eV – 6547.9 eV, which are close to that of $Mn_2O_3$, hence the oxidation state can be attributed to $Mn^{3+}$,[38–41] as shown in **Figure 4a**. Co K-edge XANES spectra of samples grown at ambient are shown along with the standards CoO and $LiCoO_2$ in **Figure 4b**. All Co K-edge XANES of the Mn–Co–N films show pre-edge at ~7710 eV, the white line peak is at ~7727 eV which lies between the absorption edges of CoO and $LiCoO_2$, indicating that the valance state of Co in our films is between +2 and +3.[42]

On increasing Co content, the absorption edge is progressively shifted to positive energy (7721.7 eV to 7723.5 eV) as shown by the arrow in the inset of **Figure 4b**. This suggests a greater degree of oxidation (that is, more $Co^{3+}$) of Co exists at high Co concentrations.[33] Some oxygen is seen through the bulk of slightly Co-rich films in RBS, as discussed in SI section 2 (**Figure S3a, b**). This is common in sputtered nitride materials during the discovery stages of research, to the point where they are sometimes considered N-rich oxynitrides, but nearly O-free variants can often be achieved after continued optimization using specialized growth techniques.[3,11,43] Because the bulk oxygen concentration is set by the background pressure of oxygen-containing species during growth, and oxygen clearly prefers to bind to Co over Mn in $MnCoN_2$, we posit that a few extra percent of Co are needed to locally achieve stoichiometric $MnCoN_2$ among some CoO that is also formed.

**Figure 4c** shows Fourier-transformed EXAFS spectra ($|\chi(R)|$) at the Mn K-edge for the films with Mn/(Mn+Co) = 0.6, 0.5, and 0.4. For all films (shown in **S5**), the intensity of both the first and second shells decreases with higher Co composition, and there is no change in the shell expansion or contraction in the first and second shell, indicating the absence of structural (crystal) changes.[44] $|\chi(R)|$ results from the Co K-edge for the same Mn/(Mn+Co) = 0.6, 0.5, 0.4 positions are shown in **Figure 4d**. EXAFS pair distance data have previously been used to compare relative fractions of metals in octahedral and tetrahedral sites in Mn- and Co-containing spinel oxides[45]. Because of the similar ionic size between Mn and Co[46], peaks arise around the same places in Mn and Co radial distribution functions for a given coordination environment. For the spinel oxides, intensity at a radial distance ($r$) of 2.5 Å was attributed to octahedral Mn and Co and intensity at $r$ = 3.0 Å was attributed to tetrahedral Mn and Co.[45] Here, we see high intensity at $r$ = 2.3 Å for Mn- and Co-edge radial distribution functions, suggesting that both Mn and Co in MnCoN$_2$ are octahedral. The peak shift to lower $r$ relative to the oxide spinel is expected because the metal-anion bond lengths from octahedrally-coordinated nitride binaries are slightly smaller than in the corresponding oxides. The inset of **Figure 4d** compares the Co $K$-edge EXAFS signal of MnCoN$_2$ to standards where the Co is in both octahedral (LiCoO$_2$ in purple dashed curve) and tetrahedral (CoAl$_2$O$_4$ in orange dashed curve) environments. The film's curve is a much closer match with octahedrally-coordinated LiCoO$_2$ standard. Altogether, this EXAFS data provides strong evidence beyond XRD that ternary MnCoN$_2$ is distinct from the combination of RS MnN and ZB CoN binary phases.

Octahedral Mn$^{3+}$ is Jahn-Teller active, but we did not see any indication of peak splitting in XRD. One explanation could be that cation disorder prevents a cooperative distortion. In this case, local distortions may still be observed as low-$r$ peak-splitting in EXAFS. We do not identify any perceivable peak splitting in the MnCoN$_2$ EXAFS data, but splitting is also not present in the EXAFS data for the Mn$_2$O$_3$ standard, which is also octahedral (bixbyite) Mn$^{3+}$. Because of this, we cannot rule out Jahn-Teller activity in MnCoN$_2$.

The XANES and EXAFS of the films grown at 150 °C, 50 °C, and ambient are presented in **Figure S6a-d**. There is a consistent ordering of absorption edges for Mn and Co in XANES (**Figure S6a and Figure S6b**). The XANES data of all films show a well-defined pre-edge feature indicating a non-centrosymmetric structure.[47] The feature is more pronounced in Co-rich compositions, where a tetrahedral environment is more likely. Since there is no indication of a crystalline non-centrosymmetric structure in XRD, this feature could be due to local-tetrahedral motifs in amorphous regions or at grain boundaries. The shift in pre-edge peak position in Mn $K$-edge is smaller than in the Co $K$-edge, indicating the change in oxidation state in Mn is smaller than in Co, in agreement with XPS analysis and from EXAFS curves. Mn $K$-edge EXAFS (**Figure S6c**) shows a slightly lower magnitude for Mn/(Mn+Co) = 0.4 for films grown at all $T_d$. The consistency of Mn–N and Mn–Mn bonding in Mn EXAFS confirms the system is stable[33] at 150 °C > $T_d$ > ambient for 0.6 > Mn/(Mn+Co) > 0.4. On other hand, in Co $K$-edge EXAFS (**Figure S6d**), the Co – Co bond intensity is lower for higher Co content (lowest for the sample grown at 150 °C). This indicates that Co tends to oxidize at higher concentrations, consistent with the XPS results.

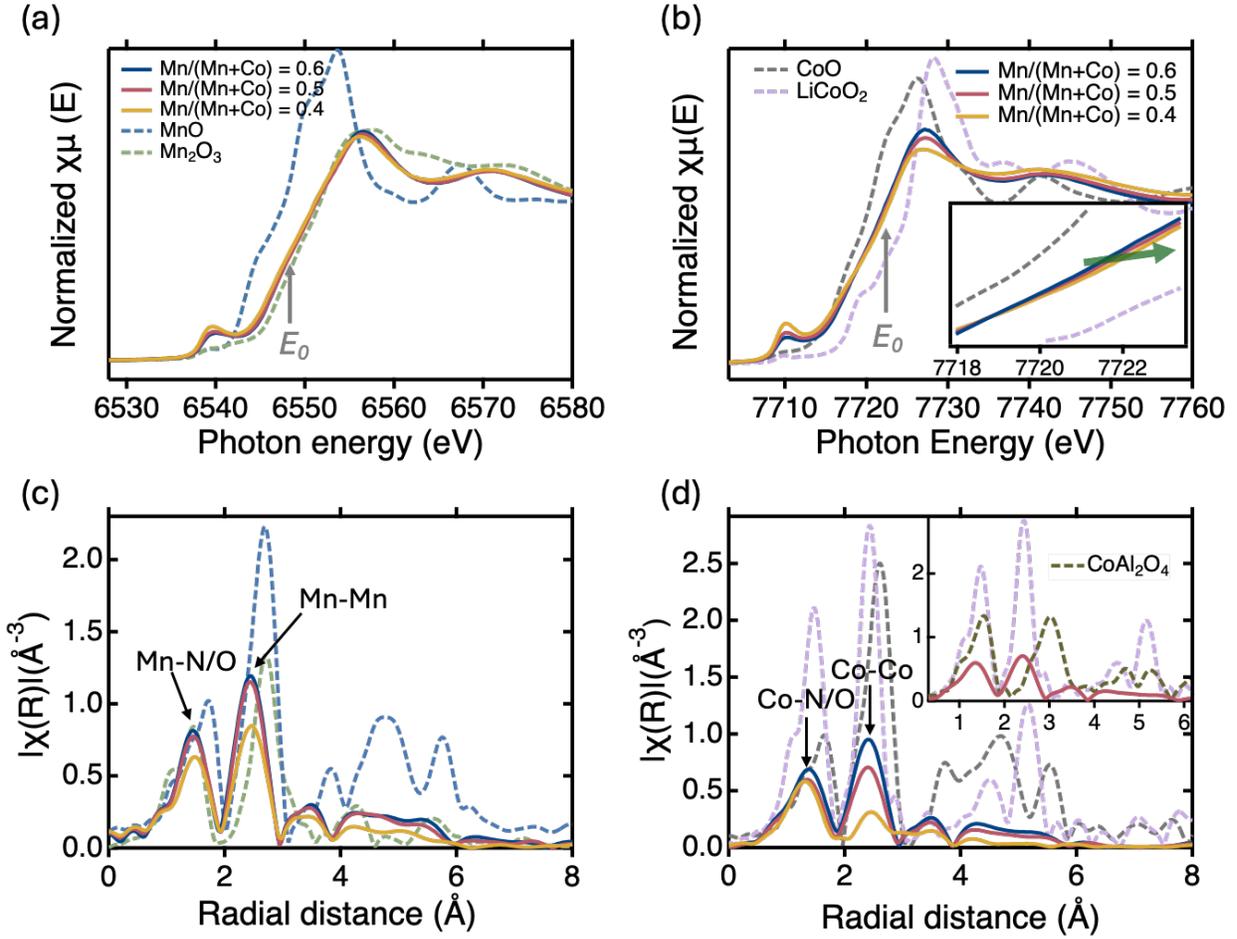

Figure 4: XAS data of films with Mn/(Mn+Co) = 0.6, 0.5, 0.4. (a) and (b) are XANES data at the Mn and Co $K$-edges. The inset of panel (b) shows the energy shift of absorption edge on increasing Co content. (c) and (d) are EXAFS data at the Mn and Co $K$-edges.

3.4. Magnetic properties

From XPS and XAS analysis, we conclude that both $Mn^{3+}$ and $Co^{3+}$ cations reside in octahedral coordination environments, with the presence of some $Co^{2+}$ that is possibly localized around O impurities. With octahedral $Mn^{3+}$, one expects four unpaired electrons in the high-spin configuration ($\mu_{eff}$ = 4.9 $\mu_B$) and two unpaired in the low-spin configuration ($\mu_{eff}$ = 2.8 $\mu_B$). For $Co^{3+}$, one expects a $S = 0$ diamagnetic low-spin configuration, whereas high-spin would have four unpaired electrons ($\mu_{eff}$ = 4.9 $\mu_B$). With $Co^{2+}$, one expects three unpaired electrons with some unquenched orbital momentum with high spin ($\mu_{eff}$ = 5.2 $\mu_B$).

DC magnetization was measured for a film with Mn/(Mn+Co) = 0.5 grown at 50 °C. **Figure 5(a)** shows the magnetization as a function of applied magnetic field, measured at temperatures ranging from 2 K – 300 K. Data are plotted after subtracting the substrate's background signal, which was measured on a comparable piece of Si. Hysteresis is observed at 2 K, and it disappears at $T = 20$ K, suggesting that a magnetic transition exists at $T < 20$ K. At 2 K, the coercive field ($H_c$) of the film is observed to be 0.4 T, and there is small net moment of $4.5 \times 10^{-5}$ emu, as illustrated by the inset in **Figure 5(a).**

Temperature dependent measurements were performed under field-cooled (FC) and zero-field-cooled (ZFC) conditions, at 1 T and 7 T. At both fields, cusps appear at 10 K (**Figure 5b).** A small bifurcation between the ZFC and FC data, and the absence of a significant shift in the ZFC peak position over a wide field range (1 T - 7 T), suggests a potential AFM transition at 10 K. However, because we observed a small hysteresis at 2 K, we hypothesize that the ground state is likely canted-antiferromagnetic order below 10 K.

We also note a change in slope in the inverse magnetization at high temperature, visible in both **Figure 5b** and a plot of *H*/moment vs. temperature (**Figure S7a**). From these data, we hypothesize that the most likely scenario is that $MnCoN_2$ is fully paramagnetic at high temperature, and short-range antiferromagnetic correlations start to disrupt ideal paramagnetic behavior below 150 K. We observed linear behavior of the moment at high temperature in a plot of the inverse moment (**Figure S7b)**, which is consistent with paramagnetic behavior, and therefore performed a Curie-Weiss fit from 170 K – 300 K. While this fit cannot yield a meaningful Curie constant as we do not know the film mass precisely enough to calculate susceptibility, a meaningful Weiss temperature can still be extracted. The Weiss temperature resulting from this fit is -49.7(3) K, supporting the assignment of AFM interactions in $MnCoN_2$.

The canted AFM ordering temperature of ~10 K is much lower than that of the reported binary Mn nitrides ($Mn_4N$ at 745 K,[48] $Mn_3N_2$ at 920 K,[49] and $MnN_{1-\delta}$ at 650 K[19]) and FM ordering of binary Co nitrides ($Co_3N$ at 615 K and $Co_{2.8}N$ at 450 K[50]). This makes our observation of a low-temperature order in $MnCoN_2$ surprising. However, cation disorder was recently found to suppress magnetic order in $MnSnN_2$ in a wurtzite-derived structure,[15] lending credibility to our hypothesis of low-temperature canted AFM order. In rocksalt $MnCoN_2$, the presence of cation disorder between diamagnetic low-spin $Co^{3+}$ and magnetic $Mn^{3+}$ would imply weak interactions, for example through magnetic dipoles, or perhaps puddles of order that do not encompass the entire sample. Further work will be required to probe the level of cation disorder in this material and explore its effect on properties. Another possible scenario for $MnCoN_2$ could be that it is AFM at all temperatures measured here like the binary Mn nitrides, although we were not able to probe this directly in this work; the bifurcation between FC and ZFC, shown in **Figure 5b**, and opening of the hysteresis loop below ~20 K (**Figure 5a,** inset) then would occur from spin canting and pinned domains that require reorientation.

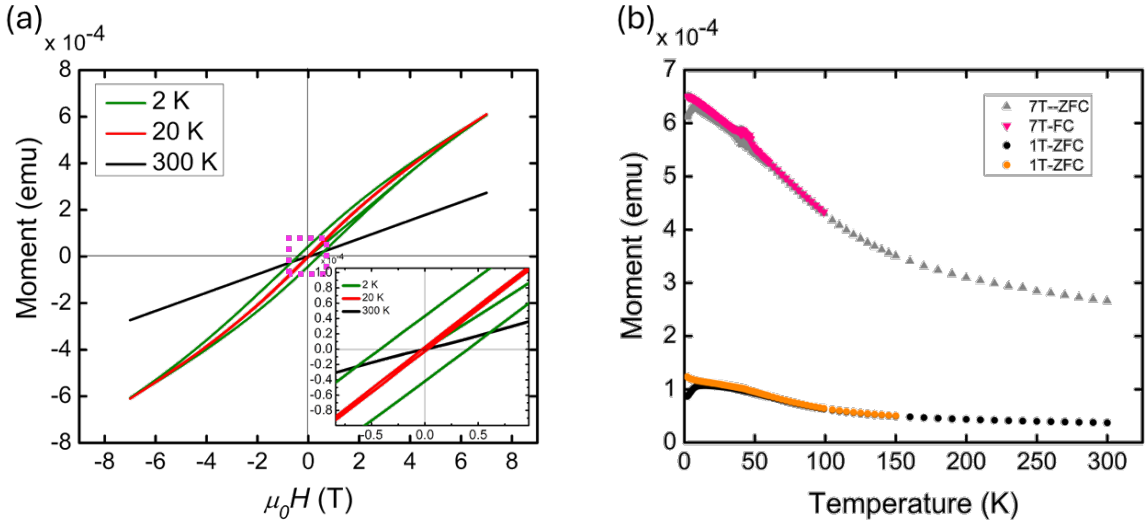

Figure 5: Magnetic properties of a film with Mn/(Mn+Co) = 0.5 grown at 50 °C. (a) Moment of the MnCoN$_2$ film as a function of applied field measured at 2 K, 20 K and 300 K. Inset is the magnified graph at lower field range. (b) Zero field cooled (ZFC) and field cooled (FC) moment as a function of temperature measured at 1 T and 7 T. Note: The FC scan was stopped once the measurement reached an identical value as the ZFC data.

4. Conclusion

A new ternary nitride material MnCoN$_2$ has been discovered in thin film form. The compound was initially predicted in a trigonal *R*3*m* space group with Mn and Co tetrahedrally bonded with four N atoms, but our calculations suggest the material is more stable in rocksalt-derived, octahedral coordination environments. Our experimental syntheses performed on different substrates confirm that MnCoN$_2$ films exhibit a cation-disordered rocksalt phase, and that the deposition temperature must be carefully controlled to achieve single-phase highly crystalline material. XPS suggests that oxygen preferentially bonds to Co on the surface of MnCoN$_2$. Further experiments using X-ray absorption spectroscopy verify octahedral coordination configuration of Co$^{3+}$ and Mn$^{3+}$ in the bulk of the film, confirming the rocksalt phase in this system. Magnetic studies show that the system possesses canted antiferromagnetic ordering with transition at ~10 K and a small net moment. In the future, the study of magnetic properties with improved film quality and under controlled cation antisite populations would be interesting. Our experimental confirmation of this new *TM$_1$*–*TM$_2$*–N material adds one more compound to the library of ternary nitrides, motivating renewed effort in new materials prediction and discovery in similar ternary spaces.

## 5. Methods

### 5.1. Synthesis

Combinatorial MnCoN$_2$ thin films were deposited using RF magnetron sputtering system at several temperature ranges of 25 – 450 ºC. The thin films were deposited by co-sputtering 3" Co (99.98%) and 3" Mn (99.99%) targets (Kurt J. Lesker Company) placed at a 120º angle to each other, thus creating a gradient in cation fluxes resulting in compositional gradient deposition of MnCoN$_2$. Prior to deposition, the chamber was evacuated to a base pressure of ~3 × 10$^{-7}$ Torr. All films were grown at a working pressure of 4 mTorr, provided by 20 sccm Ar and 5 sccm N$_2$. Most of the films were deposited on 50.8 cm x 50.8 cm pSi (100) wafers with a native oxide layer (~ 2nm), while some on 12.7 cm × 1 cm glassy carbon (HGW GmbH, Germany) and on Corning Eagle XG glass (EXG, electronically insulating and optically transparent, with a silicon back plate to transfer heat during deposition) for certain property measurements as mentioned in the text. All samples were cleaned with isopropanol and DI water followed by N$_2$ gas blow before loading into the chamber. The targets were presputtered for at least 30 mins with the shutter closed to remove surface oxide. Homogeneous binary nitrides, Mn-N and Co-N were grown by rotating the substrate during deposition for a comparative study.

### 5.2. Characterization

Experimental data for this study have been analyzed using the COMBIgor package[51], and are available publicly in the National Renewable Energy Laboratory (NREL) high-throughput experimental material database.[52,53] Thin films grown on 2" × 2" substrates were mapped as 4 X 11 points, called "libraries". A Bruker D8 X-ray diffractometer using $\theta$-$2\theta$ geometry and Cu K$\alpha$ radiation and equipped with an area detector was used to map the libraries. High resolution synchrotron grazing incidence wide angle X-ray scattering (GIWAXS) measurements were performed on select samples at beamline 11-3 at the Stanford Synchrotron Radiation Lightsource, SLAC National Accelerator Laboratory. The data were collected with a Rayonix 225 area detector at room temperature using a wavelength of λ= 0.97625 Å, a 1º or 3º incident angle, a 150 mm sample to detector distance, and a beam size of 50 μm vertical × 150 μm horizonal. Cation compositions were measured with Fischer XDV-SDD X-ray fluorescence instrument with a Rh source and a 3 mm diameter spot size. The measurements were performed at ambient temperature and pressure with an exposure time of 120 – 180 s for each measurement. Rutherford backscattering spectrometry (RBS) was used to further check the cation and anion concentration. RBS was run in a 168° backscattering configuration using a model 3S-MR10 RBS system from National Electrostatics Corporation with a 2 MeV He$^+$ beam energy. Samples were measured for a total integrated charge of 160 μC. RBS spectra were modeled with the RUPM software package.[54]

XPS measurements were performed in a Physical Electronics VersaProbe III instrument using monochromatic Al-k$\alpha$ X-rays (hν = 1486.7 eV). High-resolution XPS spectra were acquired at a 55 eV pass energy, and the binding-energy scale was were calibrated using the Au $4f_{7/2}$ (83.96 eV) and Cu $2p_{3/2}$ (932.62 eV) core levels and the Fermi edge ($E_F$ = 0.00 eV) measured on sputter-cleaned metal foils. Curve-fitting of XPS spectra was performed and fitted in Igor Pro.

X-ray absorption Spectroscopy (XAS) was performed at beamline 6-BM at the National Synchrotron Light Source (NSLS-II) at Brookhaven National Laboratory. The beamline was equipped with a paraboloid collimating mirror coated with rhodium, a monochromator utilizing (111) crystal, and a flat mirror designed for rejecting harmonic frequencies. The films on which XAS measurements were carried out were grown on glass substrates to mitigate the issue of Bragg diffraction from the crystalline substrate. Spectra of films were collected at room temperature in fluorescence mode, whereas the spectroscopic references, including CoO powder (99.5%, Sigma Aldrich), MnO powder (99.5%, Sigma Aldrich), $Mn_2O_3$ (99.9%, Sigma Aldrich), $LiCoO_2$ (99.5, sigma Aldrich), $CoAl_2O_3$ (99.5, Sigma Aldrich), were measured in transmission mode. Six scans for the Co K-edge and three scans for the Mn K-edge were collected and averaged to further improve the signal-to-noise ratio of the absorption spectra. Fourier transformed EXAFS spectra were processed with a $k^3$-weighting Hanning window (3 to 9.9 Å) and without phase correction. Data were analyzed in Athena within the Demeter software suite[55].

Magnetic properties were measured via superconducting quantum interference device (SQUID) magnetometry in a Quantum Design Magnetic Properties Measurement System (MPMS3) with the Vibrating Sample Magnetometer. The films were measured from 2 – 300 K under applied fields from -7 to +7 T. The measured $MnCoN_2$ film was an approximately 5 × 5 mm piece of a combinatorial film grown on a Si substrate. It was approximately 600 nm thick with composition Mn/(Co+Mn) = 0.5 ± 0.01 as measured by XRF. To isolate the signal of the film, a bare substrate was also measured and subtracted.

### 5.3. Calculations

In order to calculate the total energy, we employed strongly constrained and appropriately normed (SCAN) meta-GGA functional[56] in density functional theory (DFT) with the projected augmented wave (PAW) method as implemented in the Vienna *ab-initio* Simulation Package (VASP)[57–59]. We used on-site Coulomb interaction[60] of U = 2 eV for both the Mn-*d* and Co-*d* orbitals as determined in [61] for the SCAN functional. Calculations were spin polarized to account for the expected magnetic moment of Mn and Co. The soft "N_s" PAW potential was employed to allow for a reduced energy cutoff of 380 eV[62]. Further, the final energies were obtained with a Brillouin zone sampling of 4000 k-points per reciprocal atom, and the atomic forces were relaxed to below 0.01 meV/Å during the full structural relaxation.

### 6. Acknowledgements

AC36-08GO28308. Funding provided by the U.S. Department of Energy, Office of Science, Basic Energy Sciences, Division of Materials Science, through the Office of Science Funding Opportunity Announcement (FOA) Number DE- FOA-0002676: Chemical and Materials Sciences to Advance Clean-Energy Technologies and Transform Manufacturing. We thank Dr. Nicholas Strange for support with GIWAXS measurements. Use of the Stanford Synchrotron Radiation Light source, SLAC National Accelerator Laboratory, is supported by the U.S. Department of Energy, Office of Science, Office of Basic Energy Sciences under Contract No. DE-AC02-76SF00515. This research used beamline 6-BM of the National Synchrotron Light Source II, a U.S. Department of Energy (DOE) Office of Science User Facility operated for the DOE Office of Science by Brookhaven National Laboratory under Contract No. DE-SC0012704. We thank Dr. Bruce Ravel for support with XAS. The authors wish to thank the Analytical Resources Core (RRID: SCR_021758) at Colorado State University for instrument access, training, and assistance with sample analysis. The views expressed in the article do not necessarily represent the views of the DOE or the U.S. Government.

There are no conflicts to declare.

—— Supporting information ——

# Thin Film Synthesis, Structural Analysis, and Magnetic Properties of Novel Ternary Transition Metal Nitride MnCoN$_2$

Sita Dugu[1*], Rebecca W. Smaha[1], Shaham Quadir[1], Andrew Treglia[2], Shaun O'Donnell[1,2], Julia Martin[1], Sharad Mahatara[1], Glenn Teeter[1], Stephan Lany[1], James R. Neilson[2,3], Sage R. Bauers[1†]

[1] *Materials Sciences Center, National Renewable Energy Laboratory, 15013 Denver West Parkway, Golden, Colorado 80401, United States.*
[2] *Department of Chemistry, Colorado State, University, Fort Collins, Colorado 80523-1872, United States.*
[3] *School of Advanced Materials Discovery, Colorado State University, Fort Collins, Colorado 80523-1872, United States.*

\* sita.dugu@nrel.gov
† sage.bauers@nrel.gov

**Contents:**

Section S1: Structural characterization

The crystal structures of all the synthesized films were examined by laboratory X-ray diffraction (XRD) using a Bruker D8 and on selected samples with synchrotron grazing incidence wide angle X-ray scattering (GIWAXS). The samples grown at low temperature (at ambient and at 50 ºC) exhibit highly crystalline peaks with presence of rocksalt (RS) and zincblende (ZB) phases at different Mn/(Mn+Co) contents. However, the samples grown at higher temperature possess a single phase (RS), but the crystallinity is reduced with increasing temperature. Figure S1a is the heat map of laboratory XRD patterns of 44 points of the film deposited at 150 ºC. Unlike in Figure 2a, Figure S1a consists of the (111) and (200) identical to RS phase (gray curve) throughout the composition. The black and gray curves at the bottom of heatmap are ZB and RS references. Figure S1b shows the full width half maxima (FWHM) of the reflection peak (111) and (200) for the stochiometric composition; i.e, Mn/(Mn+Co) = 0.5, on the films grown at ambient, 50 ºC and 150 ºC. The FWHM increases progressively at higher deposition temperature ($T_d$), indicating the structure is less crystalline at high $T_d$.

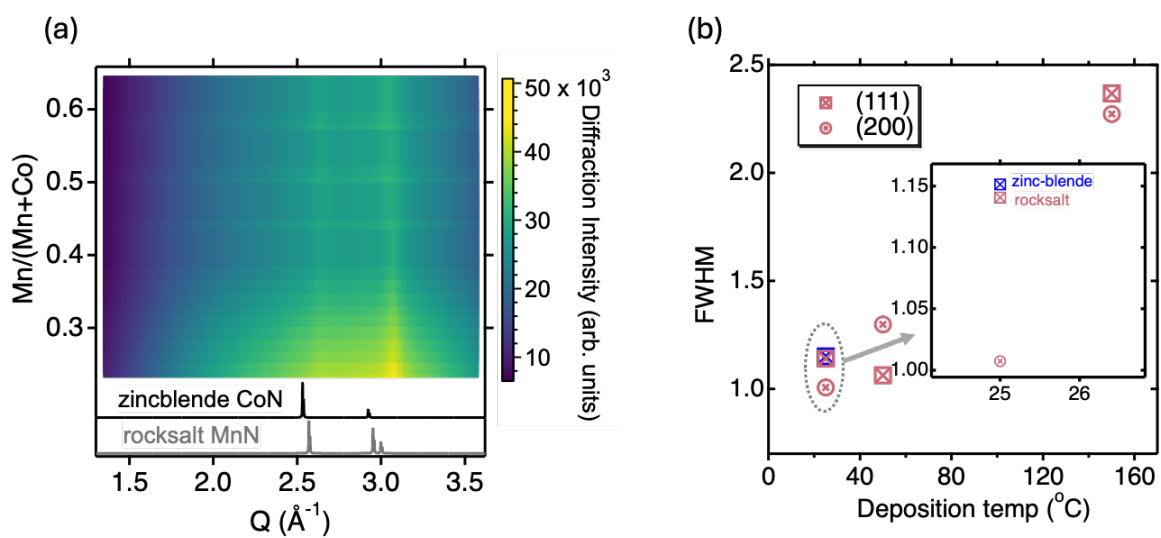

Figure S1: (a) Laboratory XRD heatmap as a function of Mn/(Mn+Co) for the library film grown at 150 ºC. Black and gray curves are reference diffraction patterns for the ZB and RS phases. (b) FWHM of the (111) and (100) reflections for the stochiometric composition. The inset shows the magnified FWHM for the ambient growths; the ZB (111) reflection is also observed at this condition. At higher $T_d$ only the RS phase is seen.

Figures S2(a-c) and S2(d-f) are GIWAXS 2D detector images of films grown on pSi and on EXG substrates with Mn/(Mn+Co) = 0.6, 0.5 and 0.4 respectively. The bright rings in the patterns indicate the presence of crystals with random orientation, whereas the bright intensity spots seen along the Bragg peak rings are due to the presence of larger individually oriented crystals, which is distinctly seen in Figure S2-b,c,e. The angle between the peaks (111) and (200) peak (in Figure S2b) is ~47º. Figure S2(a,b,c) consists of double rings near the (111) and (200) reflections, indicating the presence of both ZB and RS phases as discussed above. On comparing Figures S2b and S2e which are both stoichiometric composition, we observed double peaks in (111) and (200) in S2b but not in S2e, confirming that the films are stable as a single RS phase in EXG substrate. Also, the sharp and single rings observed in Figure S2(e-f) indicate the films are much more crystalline in EXG substrate at ambient.

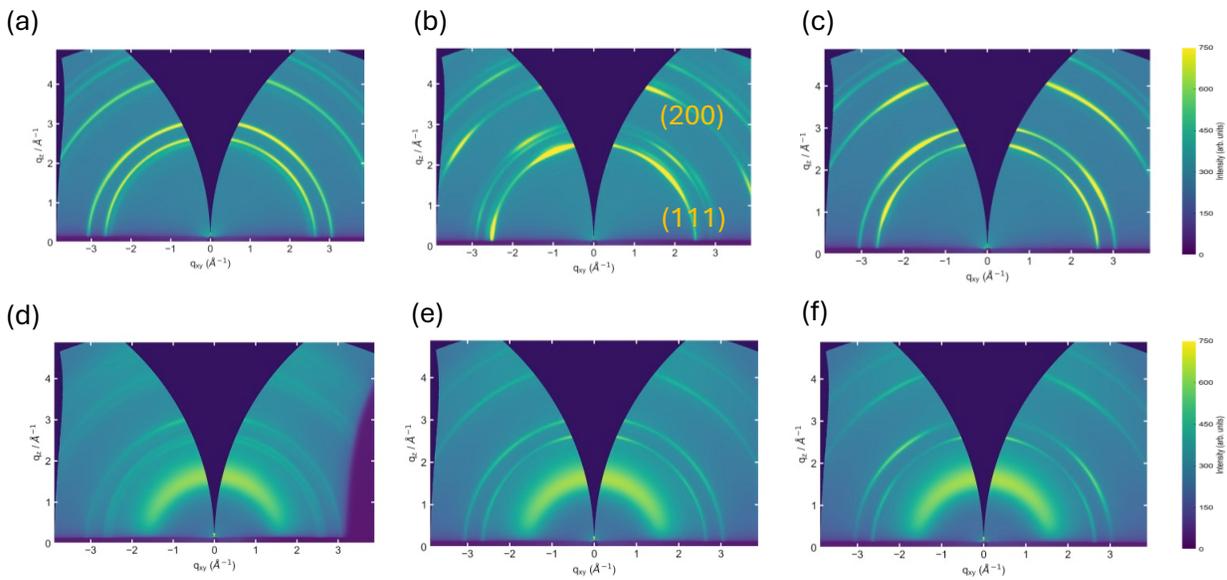

Figure S2: (a-c) and (d-f) are GIWAXS 2D detector images for Mn/(Mn+Co) = 0.6, 0.5, and 0.4 grown on pSi and EXG substrates, respectively. Both films were grown at ambient condition.

Section S2: Composition characterization

Cation concentrations are measured with X-ray fluorescence (XRF). To observe the anion (N and O) contents, RBS is carried out at six points of a film grown on glassy carbon, which helps to reduce the convolution of the film and substrate signals. Mn and Co contents were also measured by RBS. The data shows Mn:Co:N ratio ~1:1:2, with some presence of countable amount of oxygen. The RBS ratio is Mn:Co:N:O = 1:1.2:1.59:0.6. The RBS data is shown in black, and the corresponding fit in yellow as shown in figure S3a. We observed an insignificant increase in O with the increase in Mn content shown in S3b.

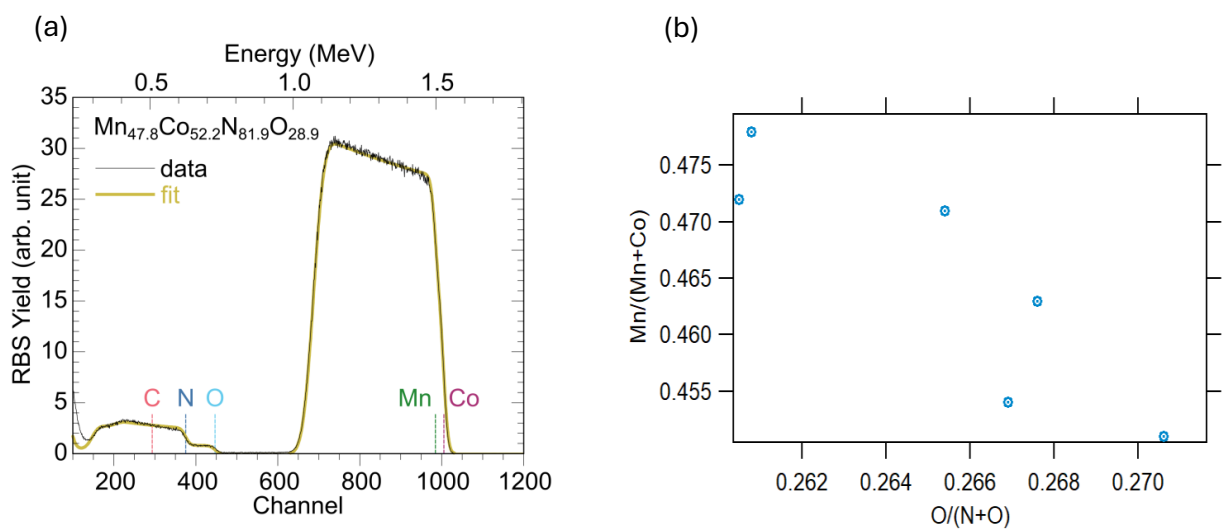

Figure S3: (a) The black solid line represents the experimental RBS spectrum, the yellow solid line represents the best fit, and other color lines indicate the main constituents within the fitted spectrum. (b) Mn/(Mn+Co) composition with oxygen concentration.

Section S3: Chemical characterization

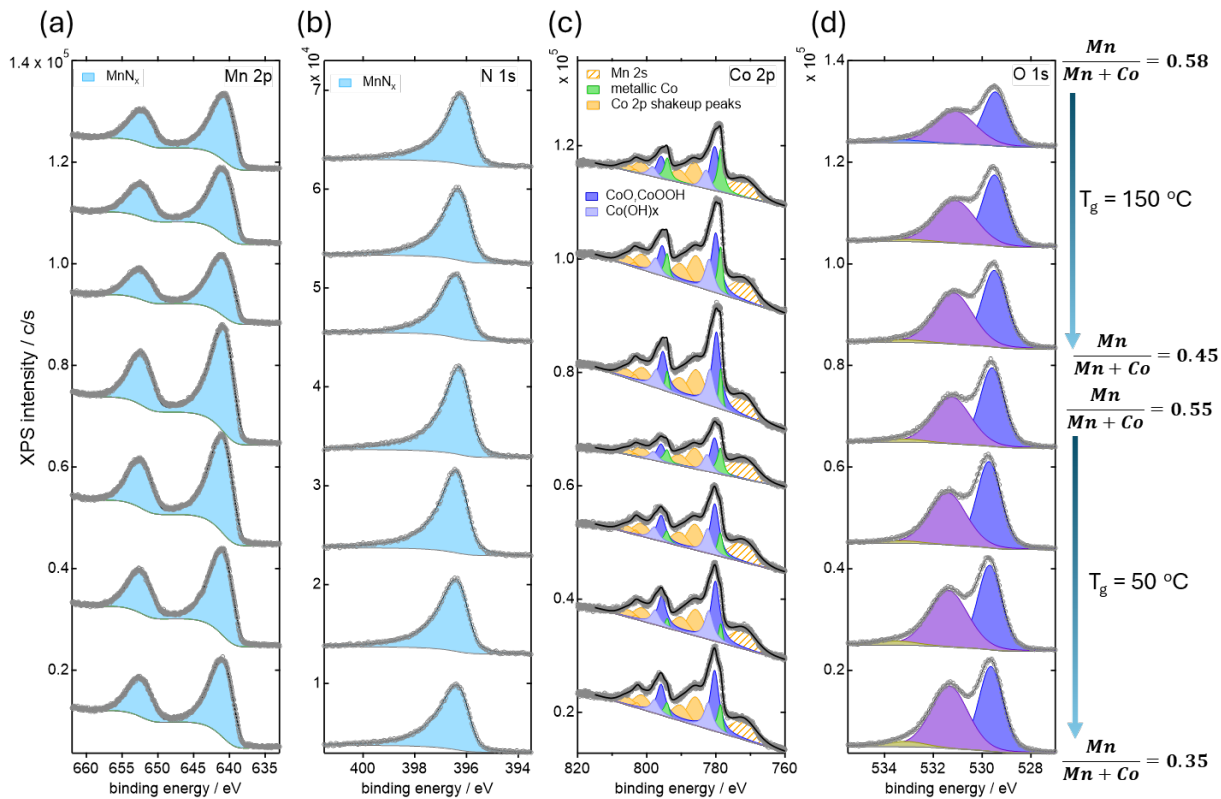

Figure S4: (a-d) XPS spectra of Mn 2p, N 1s, Co 2p, and O 1s, respectively, of films with $0.35 \leq$ Mn/(Mn+Co) $\leq 0.58$ grown at $T_d$ of 150 °C and 50 °C.

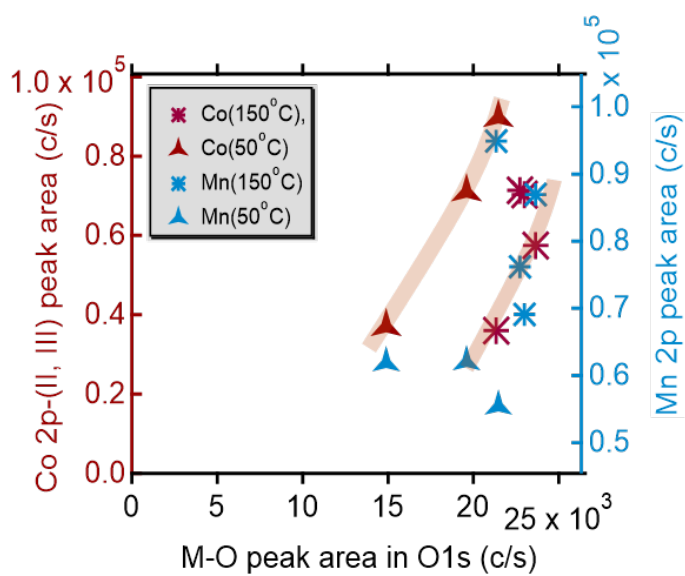

Figure S5: Cation peak area versus M – O peak area for compositional variation $0.35 <$ Mn/(Mn+Co) $< 0.58$. Co-(O/OH) positive correlation to M – O.

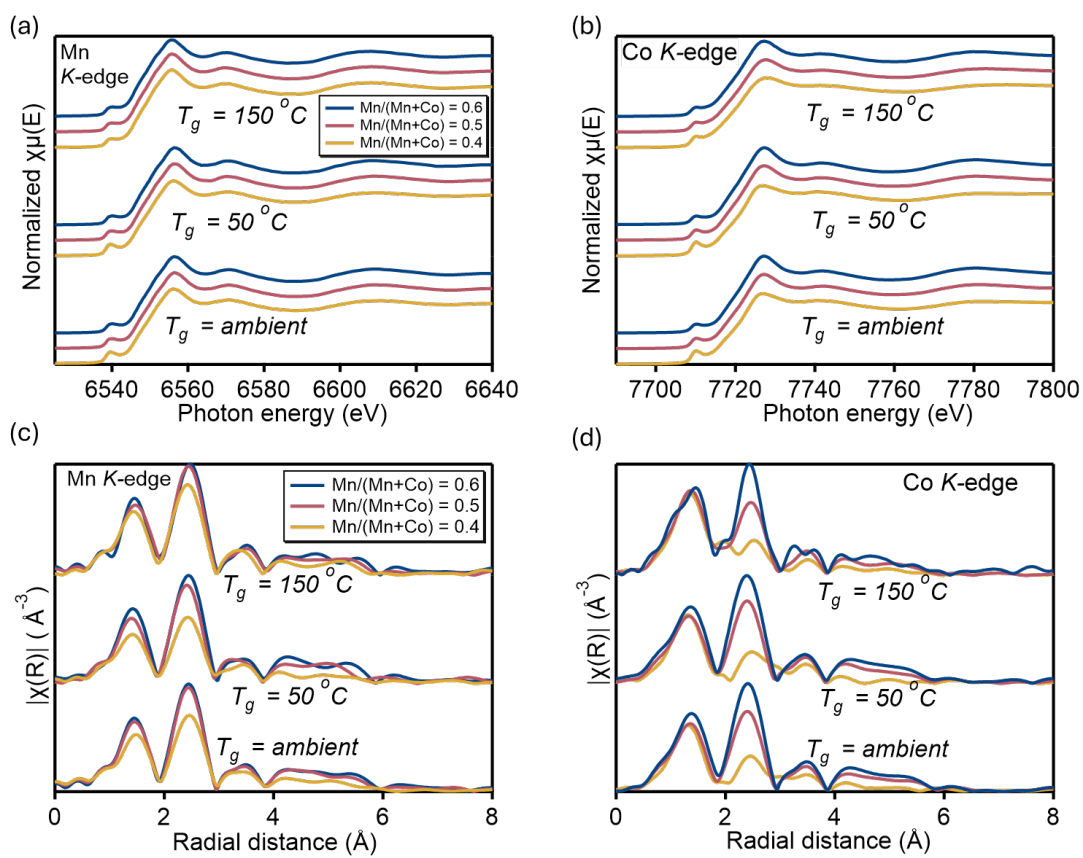

Figure S6: (a) and (b) are XANES data and (c) and (d) are EXAFS data at the Mn and Co K-edges, respectively, for the films grown at 150 °C, 50 °C, and ambient.

Section S4: Magnetic characterization

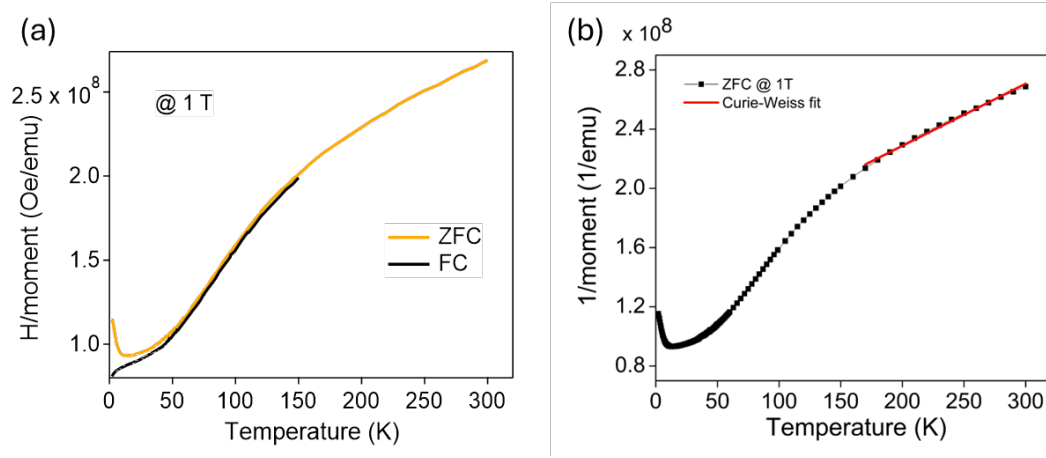

Figure S7: (a) Applied field/moment as a function of temperature measured at 1 T. (b) Inverse moment as a function of temperature measured at 1 T. The red line is a Curie-Weiss fit performed from 170 K – 300 K.